# On the Linguistic and Computational Requirements for Creating Face-to-Face Multimodal Human-Machine Interaction

João Ranhel*, Cacilda Vilela de Lima

*Abstract*— In this study, conversations between humans and avatars are linguistically, organizationally, and structurally analyzed, focusing on what is necessary for creating face-to-face multimodal interfaces for machines. We videorecorded thirty-four human-avatar interactions, performed complete linguistic microanalysis on video excerpts, and marked all the occurrences of multimodal actions and events. Statistical inferences were applied to data, allowing us to comprehend not only how often multimodal actions occur but also how multimodal events are distributed between the speaker (*emitter*) and the listener (*recipient*). We also observed the distribution of multimodal occurrences for each modality. The data show evidence that double-loop feedback is established during a face-to-face conversation. This led us to propose that knowledge from Conversation Analysis (CA), cognitive science, and Theory of Mind (ToM), among others, should be incorporated into the ones used for describing human-machine multimodal interactions. Face-to-face interfaces require an additional control layer to the multimodal *fusion* layer. This layer has to organize the flow of conversation, integrate the social context into the interaction, as well as make plans concerning 'what' and 'how' to progress on the interaction. This higher level is best understood if we incorporate insights from CA and ToM into the interface system.

## I. Introduction

Multimodal interactions between machines and humans comprise reciprocal communicative actions based on more than one communication channel, also called *modalities* (words and verbal expressions, prosody, gestures, facial expressions, eye gaze, head movement, and body posture).

Human-machine interaction is an interdisciplinary research area that is currently on the rise. Many point to Bolt's work ("Put That There") as a starting point for the study of human-machine multimodal interaction [1], [2], [3]. Research areas and subjects related to multimodal interaction include the creation of human-machine interfaces [4], [5], [6]; the recognition of facial expression and emotions [7], [8], [9], [10], [11]; machine learning in multimodal environments [12], [13], [14]; the creation of social robots [15], [16], [17], [18], [19]; mathematical and computational modeling for multimodal interactions [20], [21]; theories for creating better human-machine interfaces [4], [22], [23]; multimodal *fusion* or integration, which is related to the challenge of connecting distinct channels with different characteristics [2], [12], [24]; and affective computing [25], [26], [27], among others (see review in [28]).

One might think that talking by using natural multimodal language is the ultimate purpose of creating human-machine interfaces. Speaking naturally and being understood by the machine, even using irony and sarcasm, can seem to bring the pinnacle of efficiency to human-machine interaction. On the other hand, a machine being able to maintain the conversational flow, express opinions, demonstrate a sense of humor, and yet use convincing forms of persuasion seems to be the best we can think of for multimodal interfaces. However, this is not the reality (see [2], [3]). Many multimodal interfaces in the WIMP (window, icon, menu, pointing device) style are more efficient for certain interactive tasks than voice commands, gestures, etc. Selecting groups or areas, rotating models, pointing at something, and many other tasks are easier to do with images and pointers. Examples of the efficiency of such interfaces in the WIMP style are numerous, including CAD (computer-aided design) and IDE (integrated development environment), which are used for computer programming, video or audio editors, browsers, EDA (electronic design automation), among others.

This being so, why is there a concern with multimodal face-to-face interaction? Why do we try to create interfaces that demand so much in terms of artificial intelligence, software, hardware, and computation? Face-to-face multimodal interfaces (F2FMI) are justified for creating virtual companionship, such as that seen in *sensitive artificial listeners* [29], sociable robots, avatars for electronic games, chatterbots, and virtual attendants in kiosks or virtual stores, among other applications.

Through multimodal microanalysis, recently, one of us evaluated several interactions between humans maintaining free face-to-face conversations [30]. The conclusion was that a double feedback loop is established when two humans interact face-to-face [31]. Questions arose from this first study: (a) in qualitative terms, is it possible to notice differences in the face-to-face interaction between humans and between humans and

João Ranhel is with the Information Engineering - CECS, Universidade Federal do ABC, Av. dos Estados, 5001, S. André, CEP 09210-580, Brazil.

Correspondent author e-mail*: joao.ranhel @ ufabc.edu.br. Cacilda Vilela is with Universidade Federal do ABC, Santo André, Brazil.

avatars? Although it is laborious, it is possible to quantify the number of modal occurrences during face-to-face interaction. (b) Can the data provide evidence of the establishment of a double feedback loop? (c) How is the statistical distribution of multimodal actions/events during a free conversation between avatars and humans? The main reason to verify the existence of the double-loop feedback, as well as to determine the distributions of the multimodal events during face-to-face human-machine interactions is that designers may be aware where to spend computational efforts when creating interfaces of this nature.

In this article, we show the results of research related to the organization of face-to-face multimodal conversations between humans and intelligent multimodal interfaces. We created an F2FMI in which an avatar fully controlled by a human can talk with human volunteers without predetermined themes or tasks, a method known as Wizard of Oz (WoZ). Many articles use post-interaction forms as a research method in which users report their usability/performance experiences [32], [33]. Instead, we submit the recorded videos to scrutiny by linguistic experts. We recorded approximately 600 minutes of video and performed a linguistic microanalysis of excerpts from these interactions. We applied statistics to the data from microanalysis to understand how multimodal events are organized during an interaction.

Our article reviews the main linguistic theories needed to design human-machine interfaces for real-time multimodal interaction. We report the main concepts of conversation analysis, gestures, turn changes, facial expressions, language markers, etc., in addition to showing how double-loop feedback [31] occurs between the interactors. The experiment we realized aimed (a) to check if, in linguistical terms, multimodal interactions with avatars flow as interactions between humans; (b) to confirm the occurrence of the double-loop feedback between the interactors within a multimodal WoZ environment; and (c) to statistically obtain the distribution of multimodal actions (vocalizations, expressions, prosody, etc.) during such face-to-face interactions. We confirmed the double-loop feedback and also obtained the statistic distributions for each modality

Herein, we call the interactor who takes a turn in the conversation the *emitter*, and we call the listener during the *emitter*'s turn the *recipient*. Double-loop feedback [31] is established between them during the conversational flow, as discussed in section '5' of this article. The theoretical foundation of our research comes mainly from linguistics and conversation analysis (CA) [34], but we also borrow knowledge and concepts from the fields of gesture studies [35] and nonverbal communication [36], [37], among others. In section '2', we review the most relevant concepts from these areas related to this article. In section '3', we explain the methods used for video recording and procedures for micro-analyzing the video excerpts. In section '4', we show how the statistical inferences were carried out, as well as the results obtained.

Our research was focused on showing how both the production and the interpretation of multimodal events are distributed during a face-to-face conversation. This approach may give us an idea of the computational effort required by each multimodal unit to generate and interpret the *signs* created by the interactor. We use the term *sign* as it is used in semiotic [38]. Besides, the study helps us to understand how to organize software and hardware that integrate multimodal signs so that machines become able to maintain a smooth and free conversation. Discussion concerning these aspects, conclusions, and future work can be found in section '6'.

## II. THEORETICAL FOUNDATION

Creating interfaces for face-to-face conversation requires us to borrow theoretical knowledge from linguistic studies that address this type of interaction between humans. Despite agreeing with Oviatt and colleagues on the theoretical foundations to be applied to interfaces in general [23], we think that for face-to-face interaction interfaces it is necessary to look at the following theories, briefly explained below.

### A. Conversation Analysis

In our opinion, conversation analysis (CA) is the theory that best explains face-to-face interactions [34], [39]. CA is an approach for studying social interactions that deals with the speech in situations of casual conversation. It brings us concepts revealed by micro-level analysis of human interactions and sheds light on how people interact during a conversation. The most relevant concepts to our proposal are detailed next.

### B. Turn-taking Organization

Turning organizes social behavior in the context-situated usage of natural languages (NL). Whenever people are questioning, answering, agreeing, disagreeing, complaining, excusing, inviting, insulting, etc., they do it in turns distributed through a *turn-taking system*. The *adjacent pair* concept, which is discussed next, can only be prolific if we accept that conversation is only possible when interactors switch their role between *emitter* and *recipient* all the time.

According to Sacks, Schegloff, and Jefferson [39], a *turn* consists of vocal/verbal production, which can be composed of a single word or many sentences. This description of a turn is being questioned by researchers from gesture and nonverbal communication studies, who claim that gestural actions can also be turns. For example, a request can be carried out using distinct communicative actions, such as interrogative utterances (*Can you lend me a pen?*) or declarative statements (*I need a pen*), and so on. From the CA perspective, only such communicative actions are considered turns. However, a gesture pointing to a pen can become a request, such as "*Give me that, please*" with no words, which can cause the *recipient*'s response of delivering the pen to the requester. That is why researchers (e.g., Enfield [39], Mondada, [40]) have considered gestural movements as *turns* as well.

### C. Adjacent Pairs

One concept of CA, proposed by H. Sacks and later developed by G. Jefferson and E. Schegloff, concerns the fact that participants realize "*paired*" speeches [34], [41]. This means that what is said by an interactor restrains the other interactor's next communicative action. There are shifts in communicative actions, called *turns*. To make sense, the next turn of the other interactor should be paired with the previous ones. For instance, consider that one interactor makes an invitation to the other. To make sense and keep the conversational flow, communicative actions made in the next turn might be "to accept", "to limit the acceptance of", or "to refuse" such an invitation.

In terms of machine interface development, it is possible to use a simple solution of creating a lookup table with adjacent pairs for a given language in a given culture.

### D. Preference or Dispreference

Related to the adjacent pair concept is the organization of (*dis*)*preferred* responses [42], [43]. Let us recall the situation of an invitation. In general, when *A* invites *B* to do something, *A* expects that *B* will accept the invitation. When *B* answers positively, we say that *B*'s response is the preferred response once *B*'s response has fulfilled *A*'s expectation. Preferred actions are unnoticed; they are called unmarked actions because it is the normally expected. In other words, the *preferred* action conforms closely to the norm within a specific situation.

However, when an interactor's response disagrees with the ongoing situation, it is considered a *dispreferred* action since it does not satisfy the other interactor's expectation. According to CA, *disagreement* is closely related to (*dis*)*preferred* responses. Preferred responses are expected, as they are restricted by contexts and by cultural rules. For instance, if *A* greets *B*, it is expected that *B* will greet *A* in reply. Thus, if *B* does greet *A* in reply, this is considered a *preferred* response. However, if *B* does not greet *A* in reply, *B* could be seen as a rude interactor, i.e., such behavior could be interpreted as a *dispreferred* action for the situation.

For software creation, (*dis*)*preference* must be combined with the adjacent pair when planning the continuation of the interaction. This is justified because an agent can make an invitation out of politeness, hoping that it will be declined (preferred), while an acceptance would be *dispreferred*. To aggregate, a (*dis*)*preference* to an *adjacent pair* should be done by the software when planning a continuity for the conversation e it should not be directly associated with any lookup table.

### E. Repairs

This concept describes how interactors deal with issues that may arise when speaking, hearing, or when trying to understand the conversation (see [41], [44], [45]). A repair can be initiated by the *emitter* (called a *self-initiated repair*) or by the *recipient* (an *other-initiated repair*). It may also be performed by the *emitter* as well as by the *recipient*, resulting in four types of repairs. As in turn-taking, CA does not accept gestures or facial expressions as actions that trigger repairs; however, empirical evidence has shown that such behavior is widely used by interactors. For instance, a *recipient*'s frown can indicate to the *emitter* that the interactor has not understood the last (or the current) utterance. In this case, some researchers describe the *recipient*'s frown as an other-initiated repair (details in [46]).

Repairs, in software terms, must be done at a high-level layer that oversees whether the conversation flow is proceeding as planned. When possible, this layer must correct and return the conversation flow to what the agent intended. Repairs are closely related to what the agent interprets from the interactor's reactions, through the interpretation of real-time feedback.

### F. Intonational Unity

The concept of *intonational unity* (IU), also denoted as *prosodic units*, refers to a speech flow delivered under a single coherent intonational outline, normally marked by signs such as pauses, changes in the tone level, rhythm, volume, and the shortening or stretching of syllables or words (for details, see [47], [48]). Simply put, IU is a speech segment that occurs with a single prosodic outline. Generally, prosodic units do not correspond to syntactic units, such as phrases, clauses, or utterances.

IUs are thought to reflect different aspects of how the brain processes speech, with IUs being generated through live interaction and processing, whereas morphosyntactic units are more automated.

The development of software that inserts and interprets human prosody is a well-known subject in Artificial Intelligent in the areas of text-to-speech and speech recognition.

### G. Summing up CA and putting it into machines

From what has been explained about CA, we can see that it is a theoretical area that describes the organization of multimodal face-to-face conversations at a level above the discussions on *fusion engines*. Far beyond fusing signs with different characteristics, the set of actions and social norms explained by CA not only describes the conversation as having organized structures but also explains how/why conversation mechanisms are highly efficient. CA shows that, concurrently, the *emitter* generates multi-communicative signs, and the *recipient* produces multimodal signs as feedback, which are evaluated by the *emitter* in an attempt to verify the possible success of the conversational plan. On the other hand, the *recipient* interprets the *emitter's* signs, generating various multimodal feedback, while planning what to say as a possible conversational response in the next turn.

Consider someone trying to replicate these behaviors in a software/hardware suite. When acting as an *emitter*, the machine must generate *signs* (in semiotic terms) and interpret the human feedback. Then, it must manage the turn-taking dynamics. When turn-taking proceeds, the machine must toggle the role to '*recipient*' and then interpret the communication produced by the interactor, as well as provide feedback in some way, because humans expect them. This characterizes a double-loop feedback, which we explain elsewhere [31]. Outside of such an organization, a face-to-face conversation can be chaotic and inefficient. The terms "interpret feedback" and "interpret communication" indicate that the machine must be capable of interpreting visual and gestural signs, prosodic information, verbal sentences, and so on, have to compose them on the fly, as well as estimate this information into socio-cultural context. In other words, it has to consider the way conversation is organized within a certain culture and the context it occurs. The CA studies can provide guidelines to developers about how to organize more polite, efficient, and human-like multimodal face-to-face interactions.

*H. Theory of Mind*

To fully comprehend the aspects of face-to-face multimodal interaction (F2FMI), we also consider borrowing concepts from other areas of knowledge, specific to each modality. The most important are listed next.

The concept of "*affordance*" is used for understanding multimodal computer interfaces; however, we claim that the theory of mind [49] (ToM) is better suited for F2FMI. "Affordance theory presents a systems-theoretic view closely related to Gestalt theory. It also is a complement to Activity theory because it specifies the type of activity that users are most likely to engage in when using different types of computer interfaces. It states that people have perceptually based expectations about objects, including computer interfaces, which involve different constraints on how one can act on them to achieve goals." [23]. In contrast, "'Theory of Mind' refers to the cognitive capacity to attribute mental states to self and others. Other names for the same capacity include "commonsense psychology," "naïve psychology," "folk psychology," "mindreading" and "mentalizing" [49]."

We claim that F2FMIs are more related to causing mental states in the partner interactor than facilitating the manipulation of objects on screens, for example. That is why we need a theoretical framework for dealing with others' beliefs and the agents' own beliefs and intentions.

Machines ought to be prepared to deal with intentions, beliefs, and desires to cause consistent mental states in the partner interactor's "*mind*". The BDI (belief-desire-intention) software model may bring intuition into this framework, but delving into the BDI model is beyond the scope of this article. We believe that our communicative act is understood when the interactor makes the conversation flow naturally. In this case, we take it for granted that we caused in the interactor's *mind* mental states (interpretations) that we planned to cause.

*I. Grounding or Common Knowledge*

*Common* ground is a concept related to the set of assumptions that interaction partners share about ongoing interaction [49]. Such assumptions may concern objects and actions, as well as the interactor's understanding of the current situation and the communicative goals.

As said earlier, interactors should be able to use ToM, and they must create plans enchaining sequences of multimodal signs to pass an idea. In this sense, the set of signs and rules must be shared. Interactors who communicate face-to-face must not only share common knowledge but also foresee what such signs may cause in the other's mind. Shared knowledge is called "*grounding*" in linguistics. Despite being written in English, you would not understand a reference to Master Yoda if you did not know the Star Wars universe. This is common knowledge that we have to share if we intend to be understood.

The *grounding* concept gives us an important hint on how to build F2FMI in terms of hardware, software, and memory. Note that each interaction creates a private common ground between interactors. What someone speaks to another at a certain moment can be resumed a long time after that conversation. This indicates that we preserve memories of our interactions and we can reactivate them after a long time as a common ground between us and a specific interactor. This may be common sense, but it is aligned with the findings of other researchers (e.g., [15]). Should machines have such capability? This raises other aspects to be discussed, notably related to ethics, privacy, deception, expectations, and delusions of the users (see [19]). In this article, we intend to analyze whether human-machine interactions can occur in the same manner as between humans. In this sense, in terms of AI and development guidelines, the hint concerning *grounding* is that F2FMI should be a constant learning system and should memorize data from each interaction, as humans do.

*J. Nonverbal Communication*

The area of *nonverbal* communication (NVC) investigates how people communicate without words. NVC deals with bodily actions without considering their relation to speech and linguistic functions. In this area, bodily actions are examined from two main points of view.

First, they are considered to be indications of inner emotional states and processes (see [50]). For example, a smile indicates an individual's positive emotions, such as joy and amusement, although it may also be an index of negative feelings (see [51], [52]). Second, bodily actions are seen as communicative acts. For instance, a smile may represent an intention to affiliate, whereas a sad face may demonstrate a request for comfort (see [53] for more). Notice that a communicative act does not mean it is a linguistic act; one can communicate without using speech. Bodily actions can be communicative acts without belonging to any linguistic structure.

*K. Gestural Studies*

In a direction apart from NVC, *gesture studies* (GS) consider bodily movements as social and communicative resources, taking into consideration the relationship between bodily actions and speech. From this point of view, gestural behaviors may illustrate, emphasize, negate, or complement what is being uttered by the interactor, among other possibilities. In these terms, to a certain extent, gestures become part of a linguistic structure, and they become part of a given language. GS researchers see speech and gestural actions as a unified process, as a malleable communicational resource (e.g., [35], [54], [55], [56], [57], [58]). Moreover, the *recipient*s' bodily actions are also closely connected with what is being uttered by the speaker.

*L. Components of Multimodal Communication*

Next, we present the components of multimodal face-to-face communication considered in this research, with a brief discussion of each modality and some examples. It is out of the scope of this article to go deeper into the discussion of each modality or the intersection among modalities. What follows is used only to depict how we dealt with each modality in the interactions we videorecorded.

*M. Oral (or Vocal) Channel (or Modality)*

The oral modality is made of (a) *sounds* for generating phonemes, language markers, interjections, etc.; (b) *volume* or the intensity of sound; (c) *intonation* curves that indicate a question (ascending) or exclamation (descending), or stability; (d) *rhythm*, *pitch* or *tempo*, which indicates accelerated or slow speech; (e) *tonality*, which can be acute or severe, e.g., the louder the *emitter* speaks the more it shows irritation; (f) *sociocultural* characteristics, such as age, gender, accent, regionalism, educational level, etc.; and (g) *qualities* of the voice, which shows if the speaker is tired, excited, sad, nervous, etc. Some or all of these components are present in each event generated by this modality. The intentional combination of these components is also called *prosody* or *paralanguage*.

*N. Verbal Channel (or Modality)*

The verbal mode is made of the following. (a) *Phonemes*: these are the minimum units of meaning used to create more complex signs, such as words. Examples of phonemes are P, B, U, etc. The English language has 44 phonemes. (b) *Morphology*: these define how to join phonemes to form hierarchically more complex units. Prefixes, suffixes, infixes, and roots, among others, are all defined at the morphology level. (c) *Syntax*: this is a set of rules that dictates how to use words. Syntax rules not only dictate in which order words ought to be used to form phrases that make sense but also define the role of the words in sentences, i.e., who does the action (subject), what action is happening – or was/will be made (verb), in addition to all complements such as adjectives, adverbs, objects, etc. The grammatical description of any language is made by syntax. (d) *Semantics:* which guides how the sentence is (or should be) interpreted by the *recipient*.

*O. Gestural Channel (or Modality)*

The gestural channel is made of the following. (a) *Facial expressions:* communicate our inner state through basic emotions (e.g., fear, surprise, joy, sadness, disgust, anger, contempt, or neutral/indifference). Besides, *facial actions* (movement of lips, eyebrows, jaw, and nose) are performed during interactions to add, emphasize, deny, comment on, etc. something related to what is being communicated during the conversational flow. (b) *Eye gaze:* in a broad sense, this is used to show shared attention; for example, when someone first looks at the interactor and then turns their eyes towards an object, it can indicate that the interactor should pay attention to that object. (c) *Signature actions* or *manual gestures:* these are frequently used in human communication and have a strong impact on what we express. Hands make gestures that point to things that can complete the idea being uttered, that show rejection, nervousness, protection, and withdrawal, among others. (d) *Body actions* or *bodily postures:* actions that are also involved in human communication through movements of the head, trunk, shoulder, legs, and feet, or by the overall bodily posture, as well as by distancing between interactors.

## III. Methods

A diagram of the interactive environment is shown in Fig. 1. From the *studio* room, a person controls the facial actions, gestures, and speech of an avatar. To control the avatar's face, we use an Apple iPhone X running the Reallusion's *Live Face* app, which captures the camera images and generates a live stream of facial motion capture (mocap) data directly to a PC. The mocap stream is processed by Reallusion's *iClone 7*, which applies these data to a 3D model, transforming it into an animated humanoid. Audio is captured by lavalier microphones connected to the PC.

In the *interaction* room, a person (the interactor) talks freely with the avatar projected on the laptop screen via live meeting software. This setup is commonly referred to as the Wizard of Oz (WoZ) environment.

Three cameras are used to record the interaction. One camera captures the interactor's close-up, and another camera captures the interactor's gestures and bodily posture. To obtain synchronicity on all recorded videos, a third camera captures the avatar's close-up directly from the laptop screen.

We used Sony's Movie *Studio* for video editing and the *ELAN* software[1] to transcribe the participants' and avatars' speech and to code their bodily gestures.

*A. Participants*

Thirty-six participants (24 male/12 female) participated in this research. They filled out an open-question questionnaire carried out before interacting with the avatar. The participants ranged in age from 18 to 57 years old (M=23.61 years, SD=7.78). They were volunteers who did not receive any monetary compensation. All were Brazilian-Portuguese native speakers recruited from the Federal University of ABC, Brazil. They had different educational (e.g., undergraduate, graduate, and postgraduate) and professional backgrounds (e.g., engineer, student, teacher). Based on a five-point Likert scale, the participants described themselves as having moderate skills in technology and computer use (M=3.56 SD=1.13), and 47%

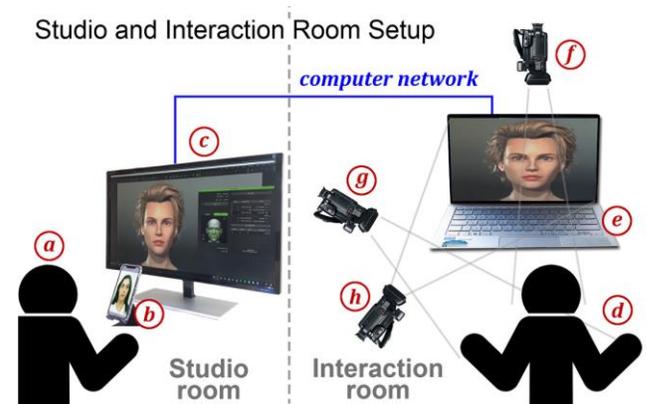

Fig. 1. Setup of the WoZ environment for recording interactions. Studio room: a person (a) uses a mobile (b) that captures the facial actions and transfers them to a PC (c) that controls a 3D-model avatar. Interaction room: a person (d) talks to the avatar on a laptop (e). Three camcorders record a close-up (f) of the interactor, the interactor's gestures and bodily postures, and (g) a close-up of the avatar (h). Audio and video are shared via live meeting software.

---

[1] *https://archive.mpi.nl/tla/elan*

declared themselves as having negligible or no experience interacting with avatars.

The participants were informed of the use of video cameras, and we obtained their permission to film them. Before recording, we explained that the purpose of our research is to study face-to-face interactions with an avatar. One-third of the participants were told *a priori* that the conversation was between them and a person, mediated by an avatar, whereas to the others, we did not explain about people controlling the avatar. When we informed *a priori* that they were talking to avatars controlled by humans, we aimed to assess whether there would be detectable differences in the use of multimodal events during the interactions, qualitatively and quantitatively. We conclude that there was not. We performed the microanalysis using the same criteria for all participants. This research was approved by the UFABC Ethics Committee - 2019/20.

*B. Multimodal Microanalysis*

More than 600 minutes were videorecorded, all of which were transcribed for other purposes. We then used 14 excerpts of ~30 seconds for further microanalysis. The fourteen excerpts analyzed were drawn by software created for this purpose, which evaluates the events on the trails generated by *ELAN* software, described below. The software randomly chooses an interaction video file, one among several minutes in length of the interview, and a starting point in 60 seconds. The only criterion for discarding a drawing excerpt is the low occurrence of multimodal events. For example, if in thirty seconds there are more than ten seconds of silence, no facial expressions, no hand gestures, nor any other multimodal event. The lengths of the excerpts were variable because we respected the beginnings and the ends of prosodic/intonational units (IU).

Microanalysis is a high-cost process that takes considerable time, attention, and discussion among analyzers. In this analytical process, we considered in detail each multimodal event produced by both the *emitter* and the *recipient*. From a linguistic and CA point of view, an event is any action with a communicating effect: a movement of the finger or head, the emission of a sound, a facial expression, and so on. Notice that *emitter*s can be either the *avatar* or the *interactor*; the same is true for *recipient*s, depending on whose turn it is to talk. Table I shows the list of multimodal events we considered in the microanalysis, including who produced the event and to which modality it belongs, as well as the identification used for the independent variable names in the spreadsheet.

The names of the variables, which also appear on the X-axis of Fig. 4, start with the letter "*E*" when the *emitter* generates the event and with "*R*" when the *recipient* does so. Thus, *Efcex* refers to the *emitter*'s facial expression, whereas *Rhdmv* refers to the *recipient*'s head movement. The variable *Silence* refers to silent moments, and *Overlap* refers to speech overlapping. *CPros* refers to moments of concomitant emission of prosody, for example, by making sound markers such as *oh, hum, wow*, etc., generated by both at the same time.

Options for annotating the occurrences of multimodal events in microanalysis include those made during a turn; however, during a turn, several intonational units (IU) may occur. A better option seems to be to use IUs as annotating markers. But, within an IU, several words, gestures, and other events may occur. The best results were obtained by using words as annotation markers. Human communication is substantially verbal. By choosing the sounds of words as a baseline for multimodal annotations, it became possible to annotate the precise moments when multimodal events occurred. It would be possible to perform even more refined microanalysis, for example, to annotate each multimodal event related to each phoneme; but doing so has not proved advantageous.

TABLE I
MODAL EVENTS CONSIDERED IN MICROANALYSIS

| item | modal event | Variable identification | |
|---|---|---|---|
| | | *Emitter* | Recipient |
| A | verbal | Evrbl | Rvrbl |
| B | prosody | Eprsd | Rprsd |
| C | facial expression | Efcex | Rfcex |
| D | eye gaze | Eeygz | Reygz |
| E | head movement | Ehdmv | Rhdmv |
| F | manual gesture | Emang | Rmang |
| G | bodily posture | Ebdpt | Rbdpt |
| H | silence | Silence | |
| I | overlap | Overlap | |
| J | concomitant prosody | CPros | |

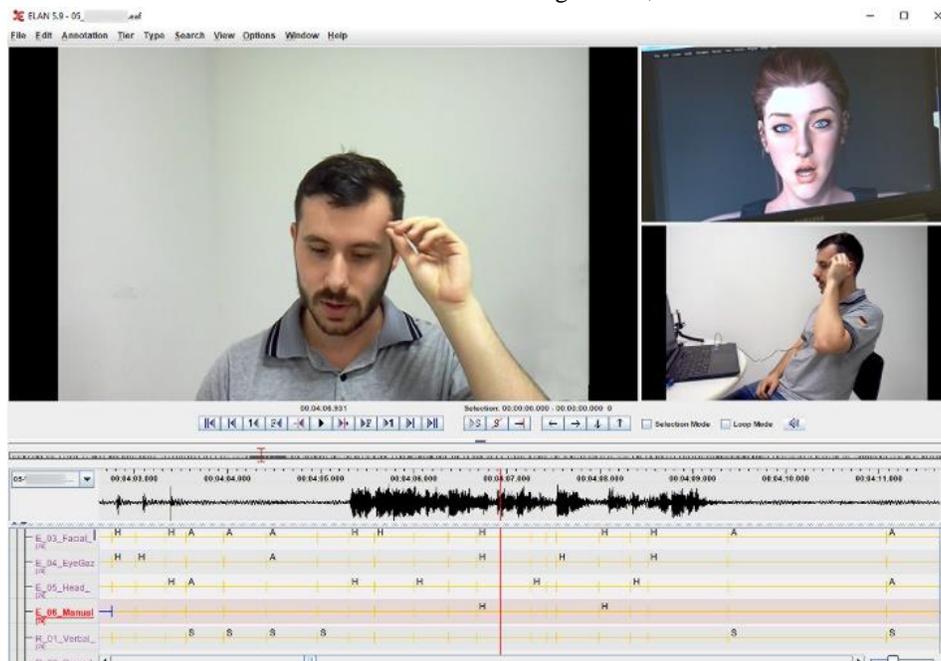

Fig. 2. ELAN interface: videos are synchronized and shown on top in separate windows. Below, users can insert "tiers" (tracks) in which the avatar and interactor's speech are transcribed and multimodal events are marked. ELAN exports such annotations as text files, in which timelines and markers are fine-grained aligned.

*C. Data Generation and Processing*

The *ELAN* software displays synchronized video windows at the top of its interface. Below these areas, users see the sound waveform. Fig. 2 shows an *ELAN* interface in which we can see the annotation tiers of a microanalysis below the soundtrack.

Below the video and soundtracks, we can insert tiers into which we can transcribe the audio or mark multimodal events of interest. *ELAN* can export such tiers as text files, keeping precise links among the video timeline and the annotation, with a resolution of milliseconds. Files with the data exported from *ELAN* were treated and statistically analyzed in R Studio.

## IV. Data Analysis and Results

We analyzed data from fourteen excerpts in which we performed microanalysis of multimodal events. In statistical terms, fourteen samples seem to be a small sample size so we used the bootstrap feature and then compared them to statistics realized without the bootstrap feature.

*A. Concurrent Modalities Distribution*

The first question we tried to answer by analyzing the data was the following: once multimodal events occur in parallel, how often do they occur in pairs, trios, etc.? We consider pairs, trios, or more groupings to be the set of multimodal events that occur simultaneously. For example, a person or the avatar can generate a vocalization and, concurrently, head movement, and facial expression, resulting in a trio of multimodal events that occurred at that particular moment. We quantified in a table the occurrence of such groups of concurrent events to apply a statistical treatment.

As we do not know either the population variance or the standard error for data of this nature, we used the Student's t-distribution with a degree of freedom (*dof*) of 13 and a 95% confidence interval ($\alpha=5\%$) for estimating such parameters; then, we calculated the confidence intervals from these samples. Fig. 3*(A)* shows the inferred populational mean for concurrent multimodal events, *dof* = 13, in which bars show the concurrency, for example, 2 = pairs, 3 = trios, etc. Also, one bar shows the silence (SL), while another shows the speech overlapping, or concomitant verbal actions (CV), and the last one shows the concurrent prosody (CP).

A low number of samples causes relatively high standard errors; so, we used the bootstrap resampling method and randomly replicated our database to 200 samples to check possible tendencies. Bootstrapping is a resampling method proposed by Bradley Efron in 1979. The Covid-19 pandemic disturbed the recording of more interactions. With the material we collected, it was possible to complete this investigation, but we performed this statistic procedure to ensure that with more data the tendency of the modal action distributions would be maintained. We used the RStudio program to raffle, with replacement, the full data files from the 14 microanalyses, and then insert the drawn files on a larger database with 200 microanalyses. We repeated the bootstrapping process ten times for checking, and in none of them did we get any statistically significant changes. The results shown in the graph refer to one of the 10 bootstrapping processes we have created, randomly chosen. Once we obtained the larger database, with 200 samples, we again estimated the confidence intervals from these samples, which can be seen in Fig. 3(B), this time we used a dof = 199. Arrows in the bars indicate the standard errors.

Note that the shape of the distributions remains practically the same if we compare Fig. 3*(A)* and 3*(B)*, but the standard errors decrease considerably on each bar in the latter one. The bar graph in Fig. 3*(B)* shows the likely result we may obtain if we use more data from other microanalyses.

Even if we chose the first analysis, with a low number of samples (only 14 microanalyses) and higher standard errors, with a 95% confidence interval, *dof* = 13, the data reveal that the mean of the population for four concurrent events falls within the interval (18.7% ± 0.52%), closely followed by three events (17.6% ± 0.83%) and five concurrent events (15.9% ± 0.84%). Pairs are also common (14.6% ± 0.75%), but silence (SL) is more frequent (15.8% ± 0.49%) than pairs or multimodal actions.

It is important to know these details because we are trying to understand the dynamics of live conversation. We intend to estimate the software and hardware requirements for dealing with face-to-face multimodal interactions. The data reveals that most of the time (≈67%), machines need to process two, three, four, or five concurrent multimodal events. This process must happen on the fly; thus, these findings may give us an idea about dimensioning the amount of data processing that a multimodal

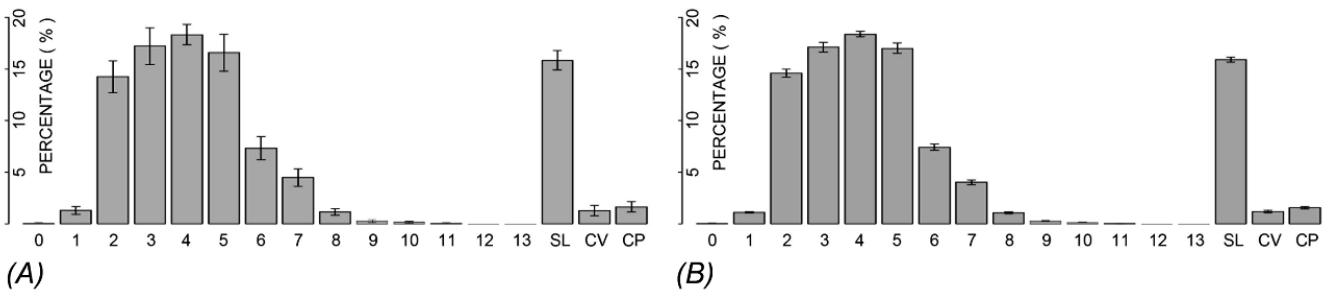

Fig. 3. Concurrent events. Percentage of pairs (2), trios (3), etc. of concurrent multimodal actions. SL=silence, CV=verbal overlap, CP=concomitant prosody. (A) Bar graph for dof = 13, α = 5%. (B) Similar bar graph after using the bootstrap method, dof = 199, α = 5%, with quite small standard errors.

interface system has to do.

In other words, we might be talking about (1) joining phonemes and identifying words, (2) analyzing the prosody of sentences, (3) processing images for identifying facial expressions, and (4) evaluating manual gestures, all of which are occurring at the same time. These four concurrent communicative actions may generate a single semantic meaning, which is why they need an integrator at this level, i.e., a *fusion* engine.

### B. Event Pairing

The next question we tried to answer was as follows: on average, how often does each multimodal event occur during interactions? Putting in another way, how are the distributions of individual occurrences of multimodal actions during a conversation? Understanding this may help us not only to dimension activities on multimodal interfaces but also to predict the tasks that need to be undertaken by the integrator or *fusion* system(s).

As discussed earlier, we use statistical inference to find the confidence intervals that described the frequency of occurrence for each event. As both the population variance and standard error of these samples are unknown, we used Student's t-test to estimate these parameters with a 95% confidence interval ($\alpha = 5\%$). The distribution for each multimodal event can be seen in Fig. 4 as a bar graph ($dof = 13$). The arrows on each bar show the standard error. The graph for 199 dof has the same shape with smaller standard errors, it is not shown.

Unsurprisingly, the two multimodal actions that prevail are prosody and verbal events. With a 95% confidence interval, the data revealed that the mean for the *emitter*'s prosody (*Eprsd*) falls within the range (13.5% ± 0.71%), larger than the second most frequent event (*Evrbl*), which is the *emitter*'s verbal events (12.99% ± 0.92%).

Let us interpret these data. First, as our sample concerns conversational interactions, as soon as there is speech production, the *emitter* controls the interaction and generates more multimodal actions, which is natural. Besides, our interactions are noticeably verbal. Every verbal sign implies prosody, but it is not reciprocal; prosody may occur without any word, e.g., in speech markers (*oh*! *wow*). Therefore, it is expected to find more prosody than verbal actions, and it is also expected to have more events coming from the *emitter* than from the *recipient*.

In addition to the predominance of *emitter*s' prosodic and verbal events, the *emitter* also generates more facial expressions (*Efcex*) than any other of the *recipient*'s actions, ranging from 10.9% (± 1.06%). In sequence, the data show that the *recipient*'s facial expressions (*Rfcex*) range from 9.3% (± 1.05%) and that the *emitter*'s head movements (*Ehdmv*) range from 9.19% (± 0.83%). Head movements are the fourth most frequent *emitter*'s action (*Ehdmv*) and the second most frequent *recipient*'s action *(Rhdmv)*. In microanalysis, all head movements are considered, without judging whether they have linguistic influence or not. Such movements usually affect what is being said, such as confirming by nodding or denying an expression with head shaking.

The reader can see that hand gestures (*Emang* and *Rmang*) are the events with the highest standard errors. It is easy to explain this outcome from the data because whereas for certain interactions, the participants had gestured a lot, in other interactions, the participants had not made a single hand movement. We believe this modality may produce the largest divergence in the distribution shown in Fig. 4, possibly due to cultural factors. Whereas some cultures use manual gestures as a natural language component, other cultures discourage or inhibit manual gestures when speaking.

## V. DISCUSSION

During microanalysis, we marked all multimodal actions in the video timeline. When both interactors performed concurrent actions, we distinguished whether the action came from the one who held the turn (*emitter*) or from the *recipient*. If face-to-face interactions are bipolar, in the style of push-to-talk radio or on half-duplex communication, then our data would only have marks for the *emitter*. When the avatar takes a turn, the human would thus be an inactive listener; vice versa, as the human takes a turn, the avatar would remain quietly unresponsive. However, face-to-face interactions are dynamic. While the *emitter* generates multimodal events, the *recipient* also responds with several multimodal actions, which mainly function as general feedback; eventually, such events generate speech overlap.

The distribution of multimodal actions seen in Fig. 4 is clear evidence that double-loop feedback occurs during the interaction. While the *emitter* produces a multimodal turn, the *recipient*'s multimodal actions can indicate understanding, or not, about what is being transmitted, and they can show whether the *recipient* agrees or disagrees with what the *emitter* utters. As soon as the *recipient* produces these signs, the *emitter* interprets them and, after taking them under consideration, can replan their turn, indicating that the *recipient*'s actions were understood as well.

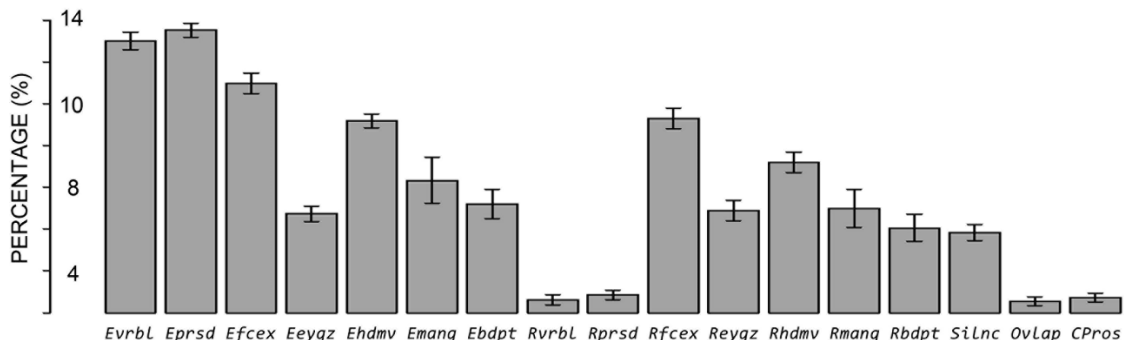

Fig. 4. Event Pairing: confidence interval showing the multimodal event occurrences (%) for the *emitter*'s and recipient's communicative actions. Please refer to the text for details. The X-axis labels are described in Table 1. (dof=199, α=5%).

On the other hand, at the time the *recipient* deciphers the set of *emitter*'s multimodal signs, the interactor makes plans for how a responsive action should be produced in a way that the conversation continues smoothly and still makes sense.

During a conversation, while the *emitter* speaks, the *recipient* plans responsive actions to be carried out as soon as the listener takes a turn. Responses are created by gathering multimodal signs that make the message either rich and intense or simple and direct but inevitably intelligible. It means to choose "*the signs*" (the multimodal actions) that may cause "*the desired effect*" on the other's mind. This requires that both interactors have a *common ground*, i.e., a shared universe of signs and rules that will be the basis for contextualizing the conversation.

That is why we believe that we need to use knowledge from the theories listed in the previous sections. As we accept the evidence of double-loop feedback, we need to understand that the *emitter* and the *recipient* follow a structure that controls the interaction. This is not just a matter of syntax, semantics, sign interpretation, contextualization, or even *sign fusion* in complex engines. Interactors expect that conversation is organized in turns, that each turn follows adjacent pairs, and that preferred responsive actions guide the construction of adjacent pairs. There is an organization to be followed when performing a face-to-face multimodal interaction.

The arguments previously exposed allow us to infer that, for F2FMI, it is necessary to create a system layer above the *sign fusion* layer, i.e., a layer to control the behavior of the "interface interactor" during the conversation so that it behaves as expected by a human during a face-to-face dialogue. This layer will take care of organizing and structuring a conversation according to protocols that humans have created and that have been used for a long time. Such a layer must be flexible enough to adapt to different cultures and individuals. To create models for this organizational layer, we suggest incorporating the knowledge of CA and ToM, among other disciplines, into the current state of the art in the multimodal interface. Designers have presented solutions that have been suggested to be quite efficient for sign interpretation modules (gestures, facial expressions, prosody, etc.), as well as solutions for *sign fusion*. However, the idea of a multimodal chat system organizer is still embryonic, and it is open to debate.

One may ask "how do the results in the paper differ or enhance the knowledge that we have from the theories discussed in Section 2?" To our knowledge, there are no articles in the scientific literature revealing data and statistics with evidence of the occurrence of a double- feedback loop in face-to-face interactions. We also hardly find studies that use statistics on the distribution of multimodal actions during face-to-face conversations, even between humans. Perhaps it is time for linguists and mathematicians to join forces to validate and solidify, quantitatively, a knowledge that until now has only been established qualitatively or perceptually.

## VI. CONCLUSIONS

According to the evidence obtained from data gathered in our research, during a multimodal interaction, interactors more frequently process three, four, or five concurrent signs. We measured the distribution of such concurrency to better dimension the amount of information processing in interfaces for face-to-face multimodal interactions. Also, we measured the distribution of occurrences among multimodal actions separately. We found that prosody, verbalizations, and facial expressions are the most frequent events, followed by head movements and manual gestures. Silence and speech overlap also occur considerably.

The distinction among the multimodal actions from the *emitter* (the one who is taking a turn) and the *recipient* (the listener) indicates the existence of double-loop feedback during the interaction. In addition to the theories currently used in multimodal interface designs, we claim that knowledge from other theories listed previously in this article must be considered when designing interfaces that aim to maintain face-to-face multimodal interaction.

The replication of this experience in other circumstances, for example, with children, teenagers, elderly people, people with different socio-cultural backgrounds, etc., should bring new insights. This may reveal new distribution curves for multimodal actions that will perhaps be different from those we obtained. We hypothesize that it may occur mainly due to cultural differences. Some cultures habitually use certain multimodal actions more strongly than others or have different forms for organizing conversations. However, for any culture, it is necessary to insert a "*conversation organizer*" into the face-to-face multimodal interfaces.

ACKNOWLEDGMENT

The authors thank all the volunteers for their time spent in laborious interaction sessions. We also thank our colleagues who kindly read our article and gave us important suggestions.

## REFERENCES

[1] R. A. Bolt, "'Put-that-there': Voice and gesture at the graphics interface," in Proceedings of the 7th SIGGRAPH 1980, Jul. 1980, pp. 262–270, DOI: 10.1145/800250.807503.

[2] M. Turk, "Multimodal interaction: A review," Pattern Recognit. Lett., vol. 36, pp. 189–195, Jan. 2014, DOI: 10.1016/j.patrec.2013.07.003.

[3] S. Oviatt, "Ten myths of multimodal interaction," Commun. ACM, vol. 42, no. 11, pp. 74–81, Nov. 1999, DOI: 10.1145/319382.319398.

[4] S. Oviatt et al., "Toward a theory of organized multimodal integration patterns during human-computer interaction," in Proc of the 5th intl conf on Multimodal interfaces - ICMI '03, 2003, p. 44, DOI: 10.1145/958432.958443.

[5] S. Oviatt, B. Schuller, P. R. Cohen, D. Sonntag, G. Potamianos, and A. Krüger, Eds., The Handbook of Multimodal-Multisensor Interfaces: Language Processing, Software, Commercialization, and Emerging Directions - Volume 3. Association for Computing Machinery, 2019.

[6] S. Oviatt, B. Schuller, P. R. Cohen, D. Sonntag, G. Potamianos, A. Krüger, Eds., The Handbook of Multimodal-Multisensor Interfaces: Foundations, User Modeling, and Common Modality Combinations - Vl1. ACM, 2017.

[7] S. K. D'Mello and J. Kory, "A review and meta-analysis of multimodal affect detection systems," ACM Computing Surveys, vl47, no. 3., pp. 1–36,2015.

[8] S. Escalera et al., "Guest Editorial: The Computational Face," IEEE Transactions on Pattern Analysis and Machine Intelligence, vol. 40, no. 11.

10[9] B. Johnston and P. de Chazal, "A review of image-based automatic facial landmark identification techniques," Eurasip Journal on Image and Video Processing, vol. 2018, no. 1. Springer International Publishing, pp. 1–23, Dec. 01, 2018, DOI: 10.1186/s13640-018-0324-4.

[10] E. A. Elliott and A. M. Jacobs, "Facial Expressions, Emotions, and Sign Languages," Front. Psychol., vol. 4, p. 115, Mar. 2013, DOI: 10.3389/fpsyg.2013.00115.

[11] N. Sebe, I. Cohen, T. Gevers, and T. S. Huang, "Multimodal approaches for emotion recognition: a survey," in Proc. SPIE 5670, Internet Imaging VI, 56, Jan. 2005, pp. 56–67, DOI: 10.1117/12.600746.

[12] T. Baltrusaitis, C. Ahuja, and L. P. Morency, "Multimodal Machine Learning: A Survey and Taxonomy," IEEE Trans. Pattern Anal. Mach. Intell., vol. 41, no. 2, pp. 423–443, Feb. 2019, DOI: 10.1109/TPAMI.2018.2798607.

[13] W. Guo, J. Wang, and S. Wang, "Deep Multimodal Representation Learning: A Survey," IEEE Access, vol. 7, pp. 63373–63394, 2019, DOI: 10.1109/ACCESS.2019.2916987.

[14] D. Ramachandram and G. W. Taylor, "Deep multimodal learning: A survey on recent advances and trends," IEEE Signal Process. Mag., vol. 34, no. 6, pp. 96–108, Nov. 2017, DOI: 10.1109/MSP.2017.2738401.

[15] J. Quintas, G. S. Martins, L. Santos, P. Menezes, and J. Dias, "Toward a context-aware human-robot interaction framework based on cognitive development," IEEE Trans. Syst. Man, Cybern. Syst., vol. 49, no. 1, pp. 227–237, Jan. 2019, DOI: 10.1109/TSMC.2018.2833384.

[16] K. Rajan and A. Saffiotti, "Towards a science of integrated AI and Robotics," Artif. Intell., vol. 247, pp. 1–9, Jun. 2017, Accessed: Nov. 04, 2017. [Online]. Available: http://linkinghub.elsevier.com/retrieve/pii/S0004370217300310.

[17] A. Clodic, E. Pacherie, R. Alami, and R. Chatila, "Key Elements for Human-Robot Joint Action," in Sociality and Normativity for Robots, Cham: Springer International Publishing, 2017, pp. 159–177.

[18] C. Tsiourti, A. Weiss, K. Wac, and M. Vincze, "Multimodal Integration of Emotional Signals from Voice, Body, and Context: Effects of (In)Congruence on Emotion Recognition and Attitudes Towards Robots," Int. J. Soc. Robot., vol. 11, no. 4, pp. 555–573, Aug. 2019, DOI: 10.1007/s12369-019-00524-z.

[19] C. Tsiourti et al., "A virtual assistive companion for older adults: Design implications for a real-world application," in Lecture Notes in Networks and Systems, vol. 15, Springer, 2018, pp. 1014–1033.

[20] S. Poria, E. Cambria, R. Bajpai, and A. Hussain, "A review of affective computing: From unimodal analysis to multimodal fusion," Inf. Fusion, vol. 37, pp. 98–125, Sep. 2017.

[21] A. Vinciarelli et al., "Open Challenges in Modelling, Analysis and Synthesis of Human Behaviour in Human–Human and Human–Machine Interactions," Cognit. Comput., vol. 7, no. 4, pp. 397–413, Aug. 2015, DOI: 10.1007/s12559-015-9326-z.

[22] The Handbook of Multimodal-Multisensor Interfaces, Volume 1: Foundations ... - Sharon Oviatt, Björn Schuller, Philip Cohen, Daniel Sonntag, Gerasimos Potamianos - Google Livros. .

[23] S. Oviatt, "Theoretical foundations of multimodal interfaces and systems," in The Handbook of Multimodal-Multisensor Interfaces: Foundations, User Modeling, and Common Modality Combinations - V1, ACM, 2017, pp. 19–50.

[24] D. Lahat, T. Adali, and C. Jutten, "Multimodal Data Fusion: An Overview of Methods, Challenges, and Prospects," Proceedings of the IEEE, vol. 103, no. 9. Institute of Electrical and Electronics Engineers Inc., pp. 1449–1477, Sep. 01, 2015, DOI: 10.1109/JPROC.2015.2460697.

[25] Z. Zeng, M. Pantic, G. I. Roisman, and T. S. Huang, "A survey of affect recognition methods: Audio, visual, and spontaneous expressions," IEEE Trans. Pattern Anal. Mach. Intell., vol. 31, no. 1, pp. 39–58, 2009, DOI: 10.1109/TPAMI.2008.52.

[26] M. Pantic, A. Pentland, A. Nijholt, and T. S. Huang, "Human computing and machine understanding of human behavior: A survey," in Lecture Notes in Comp Science, 2007, v4451, p 47–71, DOI: 10.1007/978-3-540-72348-6_3.

[27] D. Portugal, L. Santos, P. Alvito, J. Dias, G. Samaras, and E. Christodoulou, "SocialRobot: An interactive mobile robot for elderly home care," in 2015 IEEE/SICE International Symposium on System Integration, SII 2015, Feb. 2016, pp. 811–816, DOI: 10.1109/SII.2015.7405084.

[28] C. P. Janssen, S. F. Donker, D. P. Brumby, and A. L. Kun, "History and future of human-automation interaction," Int. J. Hum. Comput. Stud., vol. 131, pp. 99–107, Nov. 2019, DOI: 10.1016/j.ijhcs.2019.05.006.

[29] G. McKeown, M. Valstar, R. Cowie, M. Pantic, and M. Schroder, "The SEMAINE Database: Annotated Multimodal Records of Emotionally Colored Conversations between a Person and a Limited Agent," IEEE Trans. Affect. Comput., vol. 3, no. 1, pp. 5–17, Jan. 2012, DOI: 10.1109/T-AFFC.2011.20.

[30] C. V. de Lima, "A multimodalidade na conversa face a face em episódios de desacordo," Biblioteca Digital de Teses e Dissertações da Universidade de São Paulo, São Paulo, 2018.

[31] C. Vilela and J. Ranhel, "A framework for the multimodal joint work of turn construction in face-to-face interaction," Cogn. Syst. Res., vol. 41, pp. 99–115, Mar. 2017, DOI: 10.1016/J.COGSYS.2016.07.005.

[32] L. Clark et al., "The State of Speech in HCI: Trends, Themes and Challenges," Interact. Comput., vol. 31, no. 4, pp. 349–371, Jun. 2019, DOI: 10.1093/iwc/iwz016.

[33] A. B. Kocaballi, L. Laranjo, and E. Coiera, "Understanding and Measuring User Experience in Conversational Interfaces," Interact. Comput., vol. 31, no. 2, pp. 192–207, Mar. 2019, DOI: 10.1093/iwc/iwz015.

[34] H. Sacks, Lectures on conversation. Malden: Blackwell Publishing, 1992.

[35] A. Kendon, Gesture: visible action as utterance. Cambridge: Cambridge University Press, 2004.

[36] P. EKMAN and W. V. FRIESEN, "The Repertoire of Nonverbal Behavior: Categories, Origins, Usage, and Coding," Semiotica, vol. 1, no. 1, Jan. 1969, DOI: 10.1515/semi.1969.1.1.49.

[37] P. Ekman, Emotions Revealed. London: Weidenfeld & Nicolson, 2003.

[38] W. Nöth, "Representation in semiotics and in computer science," Semiotica, vol. 115, no. 3–4, pp. 203–213, 1997, DOI: 10.1515/semi.1997.115.3-4.203.

[39] H. Sacks, E. Schegloff, and G. Jefferson, "A simplest systematics for the organization of turn-taking for conversation," Language (Baltim)., vol. 50, no. 4, pp. 696–735, 1974.

[40] L. Mondada, "Multimodal resources for turn-taking: pointing and the emergence of possible next speakers," Discourse Stud., vol. 9, no. 2, pp. 194–225, 2007.

[41] E. Schegloff, Sequence organization in interaction: a primer in Conversation Analysis I. Cambridge: Cambridge University Press, 2007.

[42] H. Sacks, "On the preferences for agreement and contiguity in sequences in conversation," in Talk and social organization, G. Button and J. R. E. Lee, Eds. Clevedon: Multilingual Matters, 1987, pp. 54–69.

[43] N. J. Enfield, Relationship thinking: agency, enchrony, and human sociality. Oxford: Oxford University Press, 2013.

[44] E. Schegloff, "The relevance of repair for syntax-for-conversation," in Discourse and syntax, T. Givón, Ed. Academic Press, 1979, pp. 261–286.

[45] E. Schegloff, G. Jefferson, and H. Sacks, "The preference for self-correction in the organization of repair in conversation," Language (Baltim)., vol. 53, no. 2, pp. 361–382, 1977.

[46] N. J. Enfield et al., "Huh ? What ? – A first survey in 21 languages," in Conversational repair and human understanding, M. Hayashi, G. Raymond, and J. Sidnell, Eds. Cambridge University Press, 2017, pp. 343–380.

[47] J. DuBois, S. Schuetze-Coburn, D. Paolino, and S. Cumming, Discourse transcription. Santa Barbara: University of California, 1990.

[48] W. Chafe, "Intonational units," in Discourse, consciousness, and time: The flow and displacement of conscious experience in speaking and writing, W. Chafe, Ed., The University of Chicago Press, 1994, pp. 53–70.

[49] H. H. Clark and S. E. Brennan, "Grounding in communication," in Perspectives on socially shared cognition, L. B. Resnick, J. M. Levine, and S. D. Teasley, Eds., American Psychological Assoc, 1991, pp. 127–149.

[50] P. Ekman, A linguagem das emoções. Sao Paulo: Lua de Papel., 2011.

[51] P. Ekman, Telling lies: Clues to deceit in the marketplace, politics, and marriage. New York and London: W.W. Norton & Company, 1992.

[52] P. Ekman and W. Friesen, "Felt, false, and miserable smiles," J. Nonverbal Behav., vol. 6, no. 4, pp. 238–252, 1982.

[53] P. Niedenthal, S. Krauth-Gruber, and F. Ric, Psychology of emotion: interpersonal, experiential, and cognitive approach. NY: Psych. Press, 2006.

[54] M. Broth and L. Mondada, "Walking away: The embodied achievement of activity closings in mobile interaction," J. Pragmat., vol. 47, no. 1, pp. 41–58, Feb. 2013, DOI: 10.1016/J.PRAGMA.2012.11.016.

[55] A. Kendon, "Semiotic diversity in utterance production and the concept of 'language,'" Philos. Trans. R. Soc. B Biol. Sci., vol. 369, no. 1651, 2014.

[56] L. Mondada, "The local constitution of multimodal resources for social interaction," J. Pragmat., vol. 65, pp. 137–156, 2014.

[57] A. Peräkylä and J. Ruusuvuori, "Facial expression and interactional regulation of emotion," in Emotion in interaction, A. Peräkylä and M.-L. Sorjonen, Eds. Oxford: Oxford University Press, 2012, pp. 64–91.

[58] J. Streeck, "Interaction and the living body," J. Pragmat., vol. 46, no. 1, pp. 69–90, 2013.